\documentclass[12pt]{article}
\usepackage{amsfonts,amssymb,amsmath,amscd,epsf}

\newcommand{\ra}{\rightarrow}
\newcommand{\Tr}{{\rm Tr}}

\newcommand{\hc}{{\hat c}}

\newcommand{\ZZ}{{\mathbb Z}}

\newcommand{\tF}{{\tilde F}}

\newcommand{\tC}{{\tilde C}}

\newcommand{\bS}{{\bar S}}
\newcommand{\bH}{{\bar H}}
\newcommand{\bV}{{\bar V}}
\newcommand{\bz}{{\bar z}}
\newcommand{\bpsi}{{\bar\psi}}
\newcommand{\bchi}{{\bar\chi}}
\newcommand{\bc}{{\bar c}}
\newcommand{\bsigma}{{\bar\sigma}}
\newcommand{\bD}{{\bar D}}
\newcommand{\tA}{{\tilde A}}
\newcommand{\tOmega}{{\tilde\Omega}}
\newcommand{\nn}{\nonumber}
\newcommand{\cG}{\mathcal G}
\newcommand{\eps}{\epsilon}
\newcommand{\hZ}{{\hat Z}}

\newcommand{\be}{\begin{equation}}
\newcommand{\ee}{\end{equation}}

\title{Fatgraph expansion for noncritical superstrings}
\author{Anton Kapustin, Arvind Murugan\\
{\it California Instutute of Technology, Pasadena, CA 91125, U.S.A.}}

\begin{document}

\begin{titlepage}

\maketitle

\begin{abstract}
 
We study the fatgraph expansion for the Complex Matrix Quantum
Mechanics (CMQM) with a Chern-Simons coupling. In the double-scaling
limit this model is believed to describe Type 0A superstrings in
1+1 dimensions in a Ramond-Ramond electric field. With Euclidean
time compactified, we show that the
RR electric field acts as a chemical potential for 
vortices living on the Feynman diagrams of the CMQM. We interpret it as
evidence that the CMQM Feynman diagrams discretize the NSR formulation of
the noncritical Type 0A superstring.  We also study
T-duality for the CMQM diagrams and propose that a certain
complex matrix model is dual to the noncritical Type 0B
superstring.

\end{abstract}

\vspace{-5in}

\parbox{\linewidth}
{\small\hfill \shortstack{CALT-68-2495}}

\vspace{5in}

\end{titlepage}

\section{Introduction}

Recently a matrix model description was proposed for Type 0 superstrings
in 1+1 dimensions~\cite{TT,hat}. This proposal realizes an old hope that the
double-scaling limit of a matrix model is capable of providing
a nonperturbative description of (super)string theory. A particularly
intriguing aspect of Refs.~\cite{TT,hat} is that the matrix models
studied there
contain only bosonic degrees of freedom, while the dual worldsheet
theory contains fermions and is locally supersymmetric (in the NSR
formalism). Of course,
there is no paradox here, because GSO-projected vertex operators
are all bosonic. Nevertheless, it would be much more satisfying
if one could give at least a heuristic derivation of the duality on
the worldsheet level, similar to that for the 2d bosonic string. 

In the case of the 
bosonic string, the matrix model is a Hermitian Matrix Quantum Mechanics
(HMQM) with a potential 
\be
V(M)=\frac{1}{2}\Tr M^2 +\frac{\lambda}{6}\ \Tr M^3
\ee
which contains both a quadratic and a cubic piece (see Ref.~\cite{Kleb} for a concise
review.)
If we treat the cubic piece as a perturbation, the free energy of
the HMQM can be represented by a sum over trivalent fatgraphs, with Euclidean time variables living on the vertices of the diagrams. Passing to the dual graph, one gets a sum over triangulations of
compact 2d surfaces, with the Euclidean time variables living on the faces of the triangulation. In the double-scaling limit, the sum
is dominated by triangulations with many faces, and it is plausible
that the result is equivalent to 2d gravity coupled to a single
scalar field. Thus one concludes that the double-scaled HMQM describes
noncritical bosonic string in one dimension. This dimension can
be identified with the time of the HMQM. By the usual nonrigorous
argument~\cite{Polyakov}, this theory can also be thought of as a critical string
theory in 1+1 dimensions. 

In this note we discuss fatgraph expansion for the complex
Matrix Quantum Mechanics (CMQM). The basic degree of freedom
is an $N\times N$ complex matrix, and the action has $U(N)\times U(N)$ invariance. The singlet sector of this model was proposed to be
dual to the Type 0A string theory in 1+1 dimensions. The latter 
theory has
only one field-theoretic spacetime degree of freedom, the "massless tachyon" in the NS-NS sector. It also has a pair of 1-form gauge fields
in the RR sector. In 1+1 dimensions 1-forms are not dynamical, but one
can contemplate turning on a constant flux (i.e. electric field) for one or
both of them. It has been argued in Ref.~\cite{hat} that in fact
only one linear combination of the electric fields can be nonzero,
and that turning it on corresponds to working in a sector of the CMQM which is singlet with respect to the $SU(N)\times SU(N)$ subgroup, but
has a charge with respect to the $U(1)\times U(1)$ subgroup. More precisely,
the "theta-angle" of the space-time theory was identified
with the charge with respect to the difference of the two $U(1)$'s.
The theta-angle can be regarded as a chemical potential for the RR electric field.
Dually, one can say that the RR electric field is identified
with the chemical potential for the aforementioned charge of the CMQM.

An interesting question is how the RR electric field affects
the fatgraph expansion of the CMQM free energy. From the
worldsheet point of view, RR vertex operators create cuts for
worldsheet fermions. If, as in the case of the bosonic string,
one can identify fatgraphs with the discretized Type 0A
worldsheet in the NSR formalism,\footnote{This assumption may 
turn out to be false. A given string theory may admit
more than one worldsheet description: Green-Schwarz formalism
for the ten-dimensional superstring is the most famous example
of this ambiguity. It is possible that there is an alternative
to the NSR formalism for Type 0A superstring in 1+1 dimensions, and that
the fatgraphs of the CMQM discretize this alternative
worldsheet theory.}
then studying the effect of the RR field on the 
fatgraph expansion could help to ``find'' fermions in the CMQM.

A convenient way to study the nonsinglet sector of the CMQM is 
to compactify the Euclidean time and consider twisted boundary conditions
in the time direction. The RR electric field then parametrizes
the twist matrix. We will show that the Feynman diagrams of the compactified 
CMQM contain two kinds of vortices, and that the
RR electric field is the chemical potential which couples to
the difference between the net numbers of vortices of the two kinds. We interpret this
as evidence that the CMQM Feynman diagrams provide a discretization of the noncritical
Type 0A worldsheet in the NSR formalism. More specifically, we propose
that the two kinds of RR vertex operators in Type 0A
(usually denoted $(R+,R-)$ and $(R-,R+)$) correspond to two kinds of
vortices. We also argue that the scalar field living on the
Feynman diagrams of the CMQM is a sum of the worldsheet time and a worldsheet
scalar which bosonizes worldsheet fermions.

Another interesting question is the behavior of the fatgraph
expansion under T-duality. To discuss T-duality one has to work in Euclidean
signature and make the Euclidean time periodic. From the matrix-model point of
view, this amounts to studying the canonical partition function of the model.
In the bosonic case, one expects that T-duality inverts the periodicity of the Euclidean
time, and this can be seen already on the the discretized worldsheet (before taking
the double-scaling limit), provided one
excludes the contribution of vortices~\cite{GK,Kleb}. In the superstring
case, we expect T-duality to exchange Type 0A and Type 0B 
theories. Applying the dualization procedure to the fatgraphs
of the CMQM, we will find that the dual Feynman rules can be derived
from a certain complex matrix model. We argue 
that in a suitable scaling limit this model describes Type 0B
string theory and explain how this proposal is related with Refs.~\cite{TT,hat},
where the same theory was proposed to be described by an HMQM with an inverted
harmonic oscillator potential and symmetric eigenvalue distribution.

\section{CMQM and the Type 0A superstring}

The CMQM action has the following form:
\be
S=\int dt\ \Tr \left[\frac{dM^\dag}{dt}\frac{dM}{dt}-V(M^\dag M)\right].
\ee
Here $M$ is a complex $N\times N$ matrix, and the potential can be
taken to be
\be\label{pot}
V(x^2)=m^2 x^2+\frac{\lambda x^4}{2}
\ee
This theory has $U(N)\times U(N)$ symmetry:
\be
M\ra U M V^{-1},\quad U,V\in U(N).
\ee 
The diagonal $U(1)$ subgroup 
$$
U(1)_{diag}=\left\{ U=V=e^{i\alpha}\cdot {\bf 1}\right\}\subset U(N)\times U(N)
$$
acts trivially, so it is more accurate to say that the symmetry group is 
$G=(U(N)\times U(N))/U(1)_{diag}$.
We will denote the anti-diagonal $U(1)$ 
$$
\left\{ U=V^{-1}=e^{i\alpha}\cdot {\bf 1}\right\}\subset U(N)\times U(N)
$$
by $U(1)_A$. The Hilbert
space of this theory can be decomposed into irreducible representations
of $G$. If one restricts the Hamiltonian to the trivial representation,
then the theory becomes equivalent to that of $N$ non-interacting fermions moving on a plane in an external potential $V(|z|^2)$
and having zero angular momentum~\cite{hat}. Thus effectively we have non-interacting fermions moving
on a half-line. In the double-scaling limit one takes $m^2$ to be negative and tunes $m^2$ and 
$\lambda$ so that the Fermi energy is nearly zero, and the number of energy levels below
the Fermi level goes to infinity. 

The double-scaling limit is believed
to describe Type 0A superstring in a simple 1+1-dimensional background;
namely, the worldsheet theory is the $N=1$ super-Liouville theory with
$\hc=9$ plus a free scalar superfield $X$. The lowest component
of $X$ is a free scalar $X^0$, and translations of this scalar
are identified with the time translations in the CMQM. The lowest component of the
super-Liouville field $\Phi$ is a scalar $\phi$. The superpartners of $X^0$ and $\phi$ will
be denoted $\chi,\bchi$ and $\psi,\bpsi$, respectively. The worldsheet action on a flat
worldsheet is:
\be
S=\frac{1}{4\pi}\int d^2z d^2\theta \left(DX\bD X+D\Phi \bD\Phi + 2 i\mu e^{\Phi}\right).
\ee
Note that the super-Liouville interaction breaks $(-1)^{F_L}$,
where $F_L$ is the left-moving worldsheet fermion number, but preserves
$(-1)^F$, where $F=F_L+F_R$ is the total worldsheet fermion number.

The spectrum of Type 0A superstring consists of a massless tachyon $T$ in the NS-NS sector and a pair of 1-form gauge fields $C, \tC$
in the RR sector. If we denote by $F$ and $\tF$ the corresponding
2-form field-strengths, the RR vertex operators in the $(-1/2,-1/2)$
picture have the following form:
\begin{eqnarray}\label{RRvert}
V&=&F(X^0,\phi)\ c \bc\ \exp\left[-\frac{\sigma+\bsigma}{2}+\phi\right] S_+\bS_-,\nn\\
\bV&=&\tF(X^0,\phi)\ c \bc\ \exp\left[-\frac{\sigma+\bsigma}{2}+\phi\right] S_-\bS_+ .
\end{eqnarray}
Here $\sigma$ and $\bsigma$ are bosonized superconformal ghosts, and $S_\pm$ and $\bS_\pm$ are 
twist operators for left-moving 
and right-moving fermions, respectively. In the asymptotic
region, where the super-Liouville interaction is negligible,
one can bosonize the fermions $\psi,\chi$ and $\bpsi,\bchi$ into a pair of
chiral bosons $H$ and $\bH$ and write the twist operators as follows:
\be
S_\pm=e^{\pm i H/2},\quad \bS_\pm=e^{\pm i \bH/2}.
\ee
If we introduce a nonchiral $2\pi$-periodic scalar field $\xi=H+\bH$, then RR vertex operators
carry unit winding and zero momentum for $\xi$. The 0A GSO projection amounts to keeping only states
with zero momentum in the $\xi$ direction. The 0B GSO projection keeps only states with zero
winding number in the $\xi$ direction.

Super-Liouville interaction breaks continuous translational invariance
for both $\xi$ and its dual. In the language of the original
fermions this simply means that rotating $\psi$ into $\chi$ or $\bpsi$ into
$\bchi$ is not a symmetry. There is still a residual $\ZZ_2$ symmetry, generated by
$(-1)^F$. It flips the sign of all worldsheet fermions.

The super-Virasoro constraints imply that the functions $F$ and $\tF$ are constant.
Thus, in the free-field approximation, there are two zero
modes corresponding to $F$ and $\tF$. It has been argued in 
Ref.~\cite{hat} that the super-Liouville interaction enforces
$F_+:=F+\tF=0$. Thus there is only one RR zero mode: $F_-=F-\tF$. 
The model admits D0-branes which are charged with respect to $F_-$~\cite{D01,D02}.
These are super-analogues of the so-called ZZ-branes~\cite{ZZ}.
The D0-brane charge $q$ can be regarded as a theta-angle, or a ``chemical potential'', for the
RR electric field, in the sense that for fixed $q$ the space-time action contains
a term
$$
q\int d^2x\ F_-.
$$
However, unlike the theta-angle in ordinary 2d QED, $q$ is integral~\cite{hat}.

On the matrix model side, it was proposed in Ref.~\cite{hat} that nonzero $q$ corresponds
to the sector of the CMQM which is singlet under the $SU(N)\times
SU(N)$ subgroup and has $U(1)_A$ charge $2qN$. To see that $q$ is integral,
recall that both $U(1)_A$ and $SU(N)\times SU(N)$ are subgroups of $G=(U(N)\times U(N))/U(1)_{diag}$.
This implies a correlation between the $U(1)_A$ charge and the transformation properties
under the center of $SU(N)\times SU(N)$. In particular, if a representation of $G$ transforms
trivially under the $SU(N)\times SU(N)$ subgroup, then its $U(1)_A$ charge is divisible by $2N$.

One may enforce the necessary selection rule on the $U(1)_A$
charge by gauging the whole $U(N)\times U(N)$ group and
adding a Chern-Simons term with a coefficient $q$ to the action:
\be
S_{gauged}=\int dt\ \Tr \left[\frac{DM^\dag}{dt}\frac{DM}{dt}-V(M^\dag M)+iq(A-\tA)\right].
\ee
Here $A$ and $\tA$ are $U(N)\times U(N)$ gauge fields, and
$$
\frac{DM}{dt}=\frac{dM}{dt}+i(AM+M\tA).
$$
Then the path integral over $A,\tA$ and $M$ with Euclidean time period $\beta$ computes
the following quantity:
\be
Z(q)=\Tr\ P_{2qN} \exp(-\beta H_{CMQM}).
\ee
Here $P_{qN}$ is the projector onto the subspace which is singlet
under $SU(N)\times SU(N)$ and has $U(1)_A$ charge equal to $2qN$.
If our goal is to understand the effect of the RR electric field on the
fatgraphs, it is more useful to work with a ``grand canonical
ensemble'', whose partition function is given by
\be
\hZ(f)=\sum_{q=-\infty}^{+\infty} Z(q) \exp\left(-iqf\right)
\ee
The partition function $\hZ(f)$ is computed by a path integral similar to that for $Z_q$, except that
only the $SU(N)\times SU(N)$ part of the gauge field is integrated over, the $U(1)_A$ part of the
gauge field is set to zero, and one imposes quasiperiodic boundary conditions on $M$:
\be
M(\beta)=e^{2if} M(0).
\ee
Following Ref.~\cite{KKK}, we may remove the $SU(N)\times SU(N)$ gauge fields from the action by writing
$$
A=-i\Omega(t) \frac{d}{dt}\Omega^{-1}(t),\quad \tA=-i\tOmega(t) \frac{d}{dt}\tOmega^{-1}(t)
$$
for some $SU(N)$-valued functions $\Omega(t),\tOmega(t)$, satisfying $\Omega(0)=\tOmega(0)=1,$
and performing the corresponding gauge transformation.
Then $M(t)$ in the path-integral has the following boundary conditions:
\be
M(\beta)=e^{2if} \Omega(\beta) M(0) \tOmega(\beta)^{-1}.
\ee
To get $\hZ(f)$ one must integrate over the equivalence classes of gauge fields.
Since the equivalence class of a gauge field on a circle is completely characterized by the
holonomy, we simply have to integrate over $\Omega(\beta)$ and $\tOmega(\beta)$ using 
the Haar measure on $SU(N)$. This last description of $\hZ(f)$ will be taken as a starting 
point for developing the fatgraph expansion of the CMQM.

As explained above, $f$ must be identified with $\beta\int dx F_-$. The integral in the
Liouville direction diverges, and must be regularized by introducing an infrared cut-off. 
By studying the dependence of the CMQM partition function on the parameter $f$,
we will be able to see the effect of the RR electric flux on the discretized worldsheet.

An alternative to the gauged CMQM with a Chern-Simons term
is an MQM where the matrix is complex and rectangular (of size
$N\times (N+q)$)~\cite{hat}. The singlet sector of this model is
equivalent to the gauged CMQM with a Chern-Simons term. Both are
equivalent to the system of planar non-interacting fermions
with angular momentum $q$ and subject to an external potential
$V(|z|^2)$. In this paper we do not consider the complex rectangular
MQM.

\section{Fatgraph expansion for the CMQM}

Consider the Euclidean path integral with twisted boundary conditions:
$$
Z(\Omega,\tilde{\Omega},f) = {\int}_{M(\beta) = e^{2if} \Omega M(0) {\tilde{\Omega}}^\dag } \mathcal{D}M(t) \  e^{- Tr \int_{0}^{\beta} dt \left[ \partial_{t} M^\dag \partial_{t} M   +
V(M^\dag M) \right] }  
$$
where $\Omega$ and $\tilde\Omega$ are $SU(N)$ matrices. As explained in the previous section,
one can identify $f$ with the RR electric field.

To develop perturbative expansion, we take 
$$ 
V(M^\dag M) = M^\dag M + \frac{\lambda}{N} (M^\dag M)^2.
$$
Expanding the path integral in the parameter $\lambda$, we generate a Feynman diagram expansion.

The propagator for complex $M$ with twisted boundary conditions reads:
$$ 
\langle M_{ik}^{\dag}(t) M_{jl}(0)\rangle_{\beta} = \sum_{m=-\infty}^{\infty} \ e^{-|t + \beta m|} 
\ e^{2i mf} \ (\Omega^{m})_{jk} (\tilde\Omega^{- m})_{il},
$$ 
where time ordering is understood on the LHS. 

When we expand in the parameter $\lambda$, factors of $Tr(M^\dag M)^2$ correspond to vertices, and the propagators are represented by double lines to keep track of the index contractions. 
Due to the twisting, indices are not conserved along propagator edges,
although indices are conserved along an edge at each vertex as shown in Fig.~\ref{fig1}.
One needs a consistent choice of direction (indicated by small arrows) for each line in the propagator to distinguish lower and upper indices of $M$. Small arrows are
``conserved'' at the vertices as a consequence of $U(N)\times U(N)$ invariance.
One also has to specify an overall direction for each propagator (large arrows)
to distinguish the $M$ end from the $M^\dag$ end.

\begin{figure}
\begin{center}
\begin{picture}(140,60)(0,0)
\thicklines
\put(10,40){\line(1,0){100}}
\put(10,50){\line(1,0){100}}
\put(11,56){\bf i} \put(107,56){\bf l}
\put(11,27){\bf k}  \put(107,27){\bf j}
\put(60,50){\line(-2,1){10}} \put(60,50){\line(-2,-1){10}}
\put(50,40){\line(2,1){10}} \put(50,40){\line(2,-1){10}}
\thinlines
\put(80,45){\line(-1,1){12}}\put(80,45){\line(-1,-1){12}}

\put(-15,0){$e^{-|t_1-t_2+\beta m|}e^{2imf} \left(\Omega^m\right)_{jk}
\left({\tilde \Omega}^{-m}\right)_{il}$}

\end{picture}
\end{center}
\caption{Feynman rule for the propagator.}
\label{fig1}
\end{figure}

\begin{figure}
\centerline{\epsfxsize=7cm\epsfbox{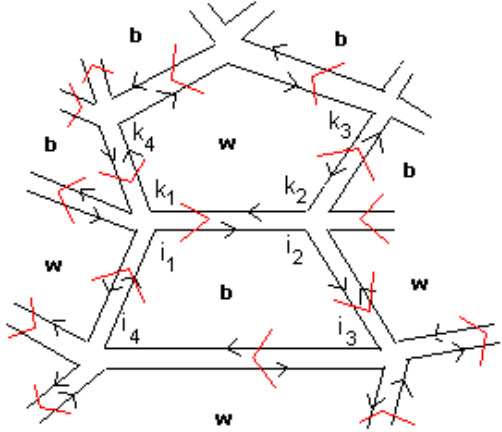}}
\caption{A bicolored Feynman diagram. The relative orientation of large and small arrows determines
the color of a face. The matrix indices are shown for some propagators and are not
conserved along the propagators due to the twisting.}
\label{fig2}
\end{figure}

Since $M^\dag$ has a non-vanishing correlator only with $M$, the large arrows around the boundary of a given face point in the same direction and define an orientation of each face. As always, small arrows determine an orientation for the entire graph which is a Riemann surface. Thus the Feynman diagrams are naturally face-bicolored: a face is colored black or white depending
on whether the orientations defined by large and small arrows along its boundary agree or disagree.
No two faces sharing an edge can have the same color, since the large arrows will point in opposite directions for them. An example of a Feynman diagram is sketched in
Fig.~\ref{fig2}. The fact that
complex matrix models lead to bicolored graphs is well-known, see Ref.~\cite{CMM} and
references therein.

Rescaling $M \rightarrow \sqrt{N} M $, the propagators acquire an additional factor of $1/N$. 
Then the weight of a given graph (with a specific coloring of faces) is
\begin{multline}\label{FD}
N^{2-2h} \lambda^{V} \prod_k \int_0^\beta dt_k \sum_{m_{ij}} \left( \prod_{white\, faces} 
\frac{1}{N} \Tr\ \tOmega^{w_{f}} \ \prod_{black\, faces} \frac{1}{N} \Tr\
\Omega^{b_{f}}
\ e^{  i (w_f - b_f) f} \right) \\
e^{ - \sum_{\langle ij\rangle } |t_i - t_j + \beta m_{ij}|} .
\end{multline}
Here $h$ is the genus of the graph, $V$ is the number of vertices, and $w_{f}$ and $b_{f}$ 
are the ``vorticities'' of the white ($w$) and black ($b$) faces: 
$w_{f} = -\sum_{\partial w} m_{ij} $ and $ b_f = \sum_{\partial b} m_{ij}$.

(Note: it is understood that in the factor $e^{ - \sum |t_i - t_j + \beta m_{ij}|}$ above, the large arrows point from $i$ to $j$ and not vice-versa. This gives precise meaning to the field 
$m_{ij}$ with respect to which the vorticities are defined. This subtlety appears to have been 
glossed over in Ref.~\cite{KKK}, where this is more pronounced due to the absence of large arrows.) 

The link variables $m_{ij}$ can be thought of as an integer-valued gauge field. Gauge transformations
are
$$
m_{ij}\ra m_{ij}+v_i-v_j,\ t_i\ra t_i-\beta v_i \quad v_i\in\ZZ\ \forall i.
$$
However, this gauge symmetry is explicitly broken by the finite integration range for $t_i$.
Therefore, we can extend the integration range for $t_i$ to the whole real axis at the
expense of gauge-fixing $m_{ij}$. The simplest gauge condition is the ``Lorenz gauge'',
which requires that at each vertex the sum of all $m_{ij}$ be zero. With this condition imposed,
$m_{ij}$ are completely determined by the face vorticities $w_f$ and $b_f$, and (for $h>0$) vorticities
around nontrivial homology cycles of the fatgraph. 

In the case of the HMQM, we have a similar expression, except that $\Omega=\tOmega$, and $f=0$.
It has been argued~\cite{GK} that diagrams with nonzero face vorticities
are dynamically suppressed in the double-scaling limit, and thus in the singlet sector we might
as well set all face vorticities to zero. This is equivalent to saying that in the double-scaling
limit it does not matter whether we integrate over the $SU(N)$-valued twist matrices $\Omega=\tOmega$ with the Haar measure, or simply set them to the identity matrix. 

In the case of the CMQM, we will assume that sectors which transform nontrivially under $SU(N)\times
SU(N)$ are separated from the singlet states by a mass gap, and therefore we can set 
$\Omega=\tOmega=1$, as in the HMQM. However, Eq.~(\ref{FD}) still depends on the face vorticities
through the RR electric flux $f$. Thus we cannot ignore vortices even in the double-scaling limit:
otherwise the effect of the RR flux would be trivial. We also see that the RR electric field
has a simple interpretation in terms of the discretized worldsheet: it is the chemical potential
for the difference of black and white vorticities. 

In the double-scaling limit the Feynman diagram expansion is dominated by diagrams with a large
number of faces. In this limit $t$ can be regarded as a scalar field living on the two-dimensional
worldsheet. The discretization of the usual free scalar field action would give
$$
\sum_{\langle ij\rangle} (t_i-t_j)^2.
$$
The action for $t$ that we are getting from the CMQM differs from this in two
respects.
First, we have to replace
$(t_i-t_j)^2$ with $|t_i-t_j|$. 
It is believed that in the double-scaling limit this does not make any difference~\cite{Kleb}.
The other difference is that we have to replace $t_i-t_j$ with
$$
t_i-t_j+\beta m_{ij},
$$
which can be thought of as a discretization of
$$
\partial_\mu t(x)-a_\mu(x)
$$
where $a_\mu$ is a gauge field on the worldsheet. The curvature of the gauge field $a_\mu$ is nonzero
only at the location of the vortices. Thus we can characterize $a_\mu$ by its holonomy
around the locations of the vortices (for $h>0$ we also have to specify holonomies
around nontrivial homology cycles). A vortex of unit strength is equivalent to 
$$
\oint a=\beta.
$$
We can eliminate $a$ by performing a multi-valued gauge transformation. This has the effect
of making the scalar $t(x)$ multivalued: it undergoes a shift by $\beta$ as one goes
around any vortex. Summation over $m_{ij}$ is equivalent to summation over all possible vortex
insertions. The meaning of this will be
discussed in Section~\ref{disc}.

\section{T-duality for the discretized superstring worldsheet}

In this section we set the RR electric field $f$ to zero, for simplicity. We also
set $\Omega=\tOmega=1$, as in the previous section. By performing T-duality
on the Feynman diagrams of CMQM, we expect to obtain the Feynman diagrams
of a dual matrix model which describes the Type 0B string compactified on the
dual circle. 

The starting point is the expression for the CMQM partition function $\hZ(0)$ derived in
the previous section:
\be\label{basicZg}
\log \hZ(0)=\sum_{G} N^{2-2h(G)} \lambda^V \prod_k \int_0^{\beta} dt_k \prod_{\langle ij\rangle}
G(t_i-t_j).
\ee
Here the first sum is over all connected four-valent fatgraphs whose faces are colored black and
white, so that as one goes around any vertex, the colors of the adjoining faces alternate. 
Each vertex is weighed by a factor $\lambda$, and the genus of the fatgraph $G$ is
denoted $h(G)$. If we replace
the quartic potential in Eq.~(\ref{pot}) by a general even polynomial of $M^\dag M$, then the 
vertices of the fatgraphs will have arbitrary even valence, but the coloring will still alternate. 

The last product in Eq.~(\ref{basicZg}) is over all edges of $G$. Each edge contributes the
following propagator:
$$
G(t_i-t_j)=\sum_{m_{ij}=-\infty}^{+\infty} \exp(-|t_i-t_j+\beta m_{ij}|).
$$
Following Refs.~\cite{GK,Kleb}, we replace the CMQM propagator with the so-called
Villain link factor:
$$
G(t_i-t_j)\ra \cG(t_i-t_j)=\sum_{m_{ij}=-\infty}^{+\infty} \exp\left(-\frac12(t_i-t_j+\beta m_{ij})^2\right).
$$
This replacement is believed not to affect the double-scaling limit and allows
for a simpler T-duality transformation. The integers $m_{ij}$ live on the
edges of the Feynman diagrams and can be regarded as a gauge field. The curl of
this gauge fields describes vortices on the worldsheet. Since we set $f=0$,
the chemical potential for vortices is zero, and we expect that vortices can be
ignored in the continuum~\cite{GK,Kleb}. Then $m_{ij}$ is constrained to be curl-free, and
on a spherical worldsheet we can write
$$
m_{ij}=v_i-v_j,
$$
for some integers $v_i$. For a worldsheet of genus $g$ one can write:
$$
m_{ij}=v_i-v_j+\eps_{ij}^A l_A,\quad v_i\in \ZZ,\ l_A\in \ZZ
$$
Here $A$ runs from $1$ to $2h(G)$ and labels arbitrarily chosen 1-cycles of the dual graph
representing the basis of the integer homology of the discretized worldsheet. 
Following Ref.~\cite{GK}, the symbol $\eps_{ij}^A$ is $\pm 1$ for an edge $\langle ij\rangle$
intersecting
the 1-cycle $A$ (the sign depends on their relative orientation), and $0$ otherwise. 
Summation over $v_i$ effectively extends the range of $t_i$ to the whole real line.
Thus the CMQM free energy becomes:
$$
\sum_{G} N^{2-2h(G)} \lambda^V \sum_{l_A}\prod_k \int_{-\infty}^{+\infty} dt_k
\prod_{\langle ij\rangle} \exp(-\frac12 (t_i-t_j+\beta \eps_{ij}^A l_A)^2).
$$
Following Refs.~\cite{GK,Kleb}, we trade integration over $t_i$ for integration over
$D_{ij}=t_i-t_j$. This requires the introduction of a Lagrange multiplier $p_I$ for each face of
the fatgraph and a Lagrange multiplier $l_a$ for each nontrivial homology cycle. Performing the
Gaussian integral over $D_{ij}$, we can express the partition
function as a sum over dual fatgraphs:
$$
\sum_{{\hat G}} N^{2-2h(G)}\lambda^F \sum_{l_a} \prod_I \int_{-\infty}^{+\infty} \frac{d p_I}{2\pi}
\prod_{\langle IJ\rangle} \exp(-\frac12 (p_I-p_J+\frac{2\pi}{\beta}\eps_{IJ}^a l_a)^2).
$$
Recall that the fatgraphs of the CMQM have bicolored faces, and as one goes around a vertex,
the color of the adjoining faces alternates. In terms of the dual graphs, this means that
vertices are colored black and white so that all the nearest neighbors of a black vertex are
white, and vice versa (this is known as a bipartite graph). Thus all edges of the dual 
fatgraph have a natural orientation (from black
to white). On each vertex of the dual fatgraph there lives a real-valued scalar $p_I$, and 
the edges are weighed with the Villain factor depending on $p_I-p_J$. Further, if the original
sum runs over four-valent fatgraphs, in the dual fatgraph all faces are quadrangles. If
we start with a more general even potential in Eq.~(\ref{pot}), then polygons with
an arbitrary even number of edges will arise. The valence of the vertices of the dual fatgraph
is not constrained. 

One can try to reproduce this sum as a Feynman diagram expansion of a matrix model. Since the
edges are oriented, it is natural to consider a complex matrix quantum mechanics of a single matrix 
$P$ with a kinetic term
$$
L_{kin}=\Tr \left[\frac{dP^\dag}{dt}\frac{dP}{dt}- P^\dag P\right].
$$
For a black vertex, all edges are outgoing, while for a white vertex, they are all incoming.
In addition, there is a symmetry which reverses the coloring. To reproduce these Feynman rules,
it is natural to try the interaction term
$$
L_{int}=\Tr\ W(P)+h.c.,
$$
where the function $W$ is a polynomial with real coefficients.
The resulting Feynman rules will weigh the vertices, rather than the faces, of the dual fatgraph, 
but in the continuum limit this should not make much of a difference. (Exactly the same
issue arises when studying T-duality for the discretized bosonic string.) 
The holomorphic potential $W(P)$ is not determined by these considerations. Note that this model
has only $U(N)$ gauge-invariance, unlike the CMQM we started from. Unfortunately, it appears 
impossible to reduce this matrix model to the eigenvalues of $P$.
Another problem is that
the potential energy is unbounded from below, and it is not clear how to give sense to
the path-integral over $P$ beyond perturbation theory. We propose an interpretation
in the next section.

\section{Discussion}\label{disc}

We have seen that the partition function of the CMQM with twisted
boundary conditions can be represented as a sum over quadrangulated
2d surfaces, where the vertices are colored black or white, so
that every black vertex is surrounded by white ones. On each
face of the quadrangulation there lives a real scalar $t$ representing
the location of the vertex of the dual graph in Euclidean time.
In addition there is a sum over vortices for $t$. These vortices
live on the vertices of the quadrangulation and therefore can be labeled
as black or white. The total vortex charge must be zero, but the
net charge of the black vortices does not have to vanish.
The RR electric field is essentially the chemical potential
for the charge of the black vortices. 

On the other hand, in the NSR formalism the vertex operator for the RR
electric field $F_-$ is the difference of $V(z,\bz)$ and $\bV(z,\bz)$ (see
Eq.~(\ref{RRvert})). Thus it is natural to interpret $V$ (resp. $\bV$) as
corresponding to the insertion of a black (resp. white) vortex on the
discretized worldsheet. This identification is supported by the following
observation. Consider the worldsheet parity operation $\Omega$. From the worldsheet viewpoint, this operation maps $V$ to $-\bV$ and $\bV$ to $-V$.
From the viewpoint of CMQM, it replaces $M$ with $M^*$~\cite{GomKap}. This has
the effect of flipping the orientation and reversing the coloring of the quadrangulation 
(black becomes white and vice
versa). In other words, it exchanges black and white vortices and also replaces all vortices
by anti-vortices. 

From the viewpoint of CMQM, the fact that the vacuum expectation values of
$F$ and $\tF$ are always opposite is obvious: this happens because the total 
vortex charge is always zero.

One interesting aspect of this identification is that the RR vertex
operator makes the scalar field $t$, living on the discretized worldsheet, multi-valued. Since
the time-like coordinate $X^0$ is univalued in the presence of RR vertex operators, this means
that the na\"ive identification of $t$ and $X^0$ cannot be correct.
Rather, it seems that one has to identify $t$ with the sum
$X^0+\frac{\beta}{2\pi} \xi$, where the scalar $\xi$ bosonizes the fermions. 
As one goes around the insertion
point of a RR vertex operator, $\xi$ under goes a shift $\xi\ra \xi+2\pi$, and thus 
$X^0+\frac{\beta}{2\pi}\xi$
will shift by $\beta$, as required. 
Note that $\xi$ does not have a continuous shift symmetry because of the super-Liouville
interaction, but $X^0$ does. This symmetry corresponds to the
time-translation invariance of the CMQM. 

As we have mentioned in the introduction, we would like to "find" the NSR fermions in the
matrix model. A weaker form of this problem is to "find" the sum over spin structures in the
fatgraph expansion of the matrix model. The identification of $t$ with $X^0+\frac{\beta}{2\pi}\xi$ 
sheds some light on this issue. In Type 0A theory, summation over spin structures is included
in the summation over all possible windings of the periodic scalar $\xi$. In the presence of
the Liouville interaction, the winding number symmetry is broken down to $\ZZ_2$. Changing
the winding of $\xi$ along some 1-cycle on the string worldsheet from even to odd is equivalent
to changing the periodicity conditions for worldsheet fermions along this cycle. From the
matrix model viewpoint, a winding for $\xi$ translates into a vortex for $t$ as one goes around
the corresponding cycle of the discretized worldsheet. 

We also discussed T-duality for the discretized Type 0A worldsheet. We found that the dual
graphs can be interpreted in terms of a perturbative expansion of a certain complex matrix
model with a $U(N)$ invariance. This model does not appear to be soluble for a generic potential.

A different matrix model for the noncritical Type 0B string was proposed in Refs.~\cite{TT,hat}.
In the double-scaling limit, it is an HMQM with a potential of the form
$$
V(M)=-M^2
$$
and eigenvalue density symmetric with respect to zero. This seems to be rather different from
what we get by T-dualizing the Type 0A matrix model. Nevertheless, it is possible that the two
models describe the same physics in an appropriate continuum limit.
Let us sketch
a plausible scenario for how this can happen. Recall that in the usual double-scaling limit
for the Hermitian matrix model one may replace the potential with an inverted harmonic oscillator
potential $V(M)=-M^2$. Although this potential is unbounded from below, one can make
sense of it by imposing a cut-off on the size of the eigenvalues of $M$. By analogy, let us
suppose that the correct scaling limit for our conjectural 0B matrix model is defined by taking $W(P)$ to be quadratic, $W(P)=\alpha P^2$. If we write $P=A+iB$, where $A$ and $B$ are Hermitian matrices,
then the Lagrangian takes the form
$$
L=\Tr\ \left[ \left(\frac{dA}{dt}\right)^2-(1-2\alpha)A^2+\left(\frac{dB}{dt}\right)^2-(1+2\alpha)B^2
\right].$$
The matrices $A$ and $B$ are decoupled.
If $\alpha>1/2,$ then $A$ is an inverted harmonic matrix oscillator, while $B$ is an ordinary
harmonic matrix oscillator. If $\alpha<-1/2$, then the roles of $A$ and $B$ are reversed. For definiteness, let us choose the first possibility. Then the partition function for $B$ does not have
any singularities and can be ignored in the continuum limit. On the other hand, the partition
function for $A$, with the cut-off imposed, has a nonanalytic behavior if one tunes $N$ appropriately.
In the singlet sector, this nonanalyticity arises when the Fermi-level of the equivalent free-fermion
model approaches zero. The continuum limit should correspond to picking the nonanalytic
terms in the partition function for $A$. We conclude that in this limit our complex matrix model
for $P$ becomes equivalent to the Hermitian matrix model with an inverted harmonic oscillator
potential. This agrees with the proposal of Refs.~\cite{TT,hat}.

\section*{Acknowledgments}

A.K. would like to thank Michael Douglas and Jaume Gomis for useful discussions. This work was supported in part by the DOE grant DE-FG03-92-ER40701.


\begin{thebibliography}{99}





\bibitem{TT} T.~Takayanagi and N.~Toumbas,
``A matrix model dual of type 0B string theory in two dimensions,''
JHEP {\bf 0307}, 064 (2003) [arXiv:hep-th/0307083].

\bibitem{hat} M.~R.~Douglas, I.~R.~Klebanov, D.~Kutasov, J.~Maldacena, E.~Martinec and N.~Seiberg,
``A new hat for the c = 1 matrix model,'' arXiv:hep-th/0307195.

\bibitem{Kleb}
I.~R.~Klebanov, ``String theory in two-dimensions,''
arXiv:hep-th/9108019.

\bibitem{Polyakov}
A.~M.~Polyakov, ``Quantum Geometry Of Bosonic Strings,''
Phys.\ Lett.\ B {\bf 103}, 207 (1981).


\bibitem{GK}
D.~J.~Gross and I.~R.~Klebanov,
``One-Dimensional String Theory On A Circle,''
Nucl.\ Phys.\ B {\bf 344}, 475 (1990).


\bibitem{D01}
T.~Fukuda and K.~Hosomichi,
``Super Liouville theory with boundary,''
Nucl.\ Phys.\ B {\bf 635}, 215 (2002)
[arXiv:hep-th/0202032].

\bibitem{D02}
C.~Ahn, C.~Rim and M.~Stanishkov,
``Exact one-point function of N = 1 super-Liouville theory with boundary,''
Nucl.\ Phys.\ B {\bf 636}, 497 (2002)
[arXiv:hep-th/0202043].


\bibitem{ZZ}
A.~B.~Zamolodchikov and A.~B.~Zamolodchikov,
``Liouville field theory on a pseudosphere,''
arXiv:hep-th/0101152.

\bibitem{KKK}
V.~Kazakov, I.~K.~Kostov and D.~Kutasov, ``A matrix model for the two-dimensional black hole,''
Nucl.\ Phys.\ B {\bf 622}, 141 (2002) [arXiv:hep-th/0101011].


\bibitem{CMM}
P.~Di Francesco,
``Rectangular Matrix Models and Combinatorics of Colored Graphs,''
Nucl.\ Phys.\ B {\bf 648}, 461 (2003)
[arXiv:cond-mat/0208037].


\bibitem{GomKap}
J.~Gomis and A.~Kapustin,
``Two-dimensional unoriented strings and matrix models,''
arXiv:hep-th/0310195.













\end{thebibliography}
\end{document}